\documentclass[aps,amsfonts,amsmath,prd,preprint,nofootinbib]{revtex4-2}
\usepackage{epsfig,bbm,cancel}
\usepackage{slashed,xcolor}
\usepackage[normalem]{ulem} 
\usepackage{amssymb}

\begin{document}

\title{BPS Skyrmions of Generalized Skyrme Model in Higher Dimensions}

\author{Emir Syahreza Fadhilla$^{1,2}$}
\email{emirsyahreza@students.itb.ac.id}
\author{Bobby Eka Gunara$^1$}
\email{bobby@fi.itb.ac.id (Corresponding author)}
\author{Ardian Nata Atmaja$^2$}
\email{ardi002@brin.go.id}

\affiliation{$^1$Theoretical High Energy Physics  Research Division, Institut Teknologi Bandung,
Jl. Ganesha 10 Bandung 40132, Indonesia.}
\affiliation{$^2$ Research Center for Quantum Physics, National Research and Innovation Agency (BRIN),
Kompleks PUSPIPTEK Serpong, Tangerang 15310, Indonesia.}

\begin{abstract}
 In this work we consider the higher dimensional Skyrme model, with spatial dimension $d > 3$, focusing on its BPS submodels and their corresponding features. To accommodate the cases with a higher topological degree, \(B\geq 1\), a modified generalized hedgehog ansatz is used where we assign an integer \(n_i\) for each rotational plane, resulting in a topological degree that proportional to product of these integers. It is found via BPS Lagrangian method that there are only two possible BPS submodels for this spherically symmetric ansatz which shall be called as BPS Skyrme model and scale-invariant model. The properties of the higher dimensional version of both submodels are studied and it is found that the BPS Skyrmions with \(B\geq1\) exist in the first submodel but there is only \(B=1\) BPS Skyrmion in the second submodel. We also study the higher dimensional version of self-duality conditions in terms of strain tensor eigenvalues and find that, in general, the scale-invariant model has a stronger self-duality condition than the BPS Skyrme model.
\end{abstract}

\maketitle
\thispagestyle{empty}
\setcounter{page}{1}
\tableofcontents

\section{Introduction}
\label{sec:intro}

The Skyrme model was initially designed to be a model for strongly interacting particles. The construction of this model is based on the non-linear sigma model which is slightly modified by the addition of a higher-order term known as the Skyrme term \cite{Skyrme:1961vq, Skyrme:1962vh}. This modification is necessary because in the flat four-dimensional spacetime the two terms scale in opposite ways under coordinate scaling \(\Vec{x}\rightarrow \mu\Vec{x}\) which means that the solution of the Skyrme model field equations is stable under the conditions of Derrick-type stability \cite{Manton:2004tk, Derrick:1964ww}. The Skyrme model, although eventually failed to become the general theory for strong interactions, has many uses, one of which is to be an effectively good model for strong interactions in low energy regions. This low energy scheme is quite interesting to be studied extensively because there is no scheme through QCD that can predict hadronic phenomena in low energy regions \cite{Dothan:1986ae}. The growing interest in the Skyrme model emerged after the discovery of the correspondence on large \(N_c\) limit between such effective field theory in low energy regime and the general formulation of the QCD \cite{tHooft:1973alw, Witten:1983tx}.

Although Skyrme formulated this model as a model for strong interactions, currently Skyrmion is a phenomenon that can be encountered in many other non-linear systems, especially at low spatial dimensions. Examples of skyrmion phenomena that can be found in solids varied from the superfluid polariton \cite{Donati14926}, Bose-Einstein ferromagnets \cite{Khawaja:2001zz}, antiferromagnets \cite{Baskaran:2011wg}, magnetic thin films \cite{Kiselev_2011}, to chiral nematic liquid crystal \cite{Fukuda_2011}. In addition, the phenomenon of a skyrmion is also quite common in optics such as evanescent waves \cite{Tsesses993}. This raises a new challenge in the generalization of the Skyrme model in any spacetime dimension. In fact, we have an exotic matter in higher spacetime dimensions that is constructed in the same manner, namely skyrmion string that belongs to the \(D\)-brane solitons \cite{Gudnason:2014uha, Prasetyo:2017rij}. The Skyrmion string is a Skyrmion-like string which ends on domain walls since the domain walls are located on the potential term's vacua and supported by \(\pi_{d-1}(S^{d-1})=\mathbb{Z}\)\footnote{It is conventional to have potential terms with the vacuum expectation value of the skyrmion field minimizes the potential, i.e. become the vacua of the potential terms.}. The original construction of this skyrmion can only be carried out up to \(1+6\) dimensional spacetime since the Skyrme action contains only up to sextic order term. A similar object can also be found in the holography theory, known as Skyrme black brane where a Skyrme model coupled with gravity in asymptotically five-dimensional AdS spacetime \cite{Cartwright:2020yoc}. Another brane-like object in \(1+6\) dimensions that has a skyrmionic solution is the Skyrme brane which has a slightly different origin from the previously mentioned skyrmion string \cite{Blanco-Pillado:2008coy}. A skyrme brane constructed from a skyrme action up to quartic derivative term (also known as the Skyrme term) governing the dynamics of the Skyrme field as a map from three-dimensional submanifold of \(1+6\) dimensional spacetime (the \(3\)-brane) to a \(S^3\) manifold represented as \(SU(2)\)-valued field. It is a gravitating model with seven-dimensional Einstein-Skyrme action possessing a spherically symmetric 3-brane along the transverse directions in the bulk. In contrast with the previous D-brane soliton that lives on a completely flat higher-dimensional spacetime.

To construct such a higher dimensional object, there is a problem with the Derrick's stability conditions because the higher the spacetime dimensions, the more higher-order terms must be added. The consistent formulation for this high-order Skyrme model has been carried out in \cite{Gudnason_2017} and it is found in this way that terms consistent with the earlier Skyrme formulation can only be carried out up to sextic order for four-dimensional space-time. This generalized model in four-dimensional spacetime with sextic term have been studied in \cite{Adam:2010fg,Adam:2010zz,Adam:2020iye,Adam:2020yfv}, and the same sextic model have also been studied on five-dimensional spacetime in \cite{Brihaye:2017wqa}. One can also construct a higher-order Skyrme model through Sutcliffe truncation on quartic Yang-Mills action in an eight-dimensional spacetime by removing all vector meson terms from the gauge field expansion, leading to a Skyrme model that obeys Derrick's stability condition \cite{Nakamula:2016wwv}. It is known that this procedure can produce a higher-order Skyrme model in spacetime dimensions of multiple of four by applying Sutcliffe truncation on a \(4k\)-th order Yang-Mills action. Another approach for the extension of the Skyrme model in higher dimensions can be done by adding higher-order terms similar to the one demonstrated in \cite{Gudnason_2017} up to \(2d\)-th order term which is proportional to the squared norm of topological current and this model is originally proposed as an extension of \(O(d+1)\) sigma model \cite{Arthur:1996ia}. In this work, we start our discussion by referring to this \(O(d+1)\) sigma model construction and motivate it with the generalization of Manton's construction of static four-dimensional Skyrme model through invariants of strain tensor for higher-dimensional cases in section \ref{sect:Model}. Here we use the strain tensor eigenvalue combinations as the representation of higher-order Skyrme terms which have been proposed in \cite{manton1987} to describe the geometrical properties of four-dimensional spacetime Skyrmion as a harmonic map of a flat spatial manifold to a three-dimensional sphere manifold.

As the number of terms grows for a higher dimensional model, it is of interest to find the term which has special qualitative features. Some examples of this approach can be found in \cite{Brihaye:2017wqa} which shows that omitting quartic term does not give any significant qualitative feature in the five-dimensional spacetime model and in \cite{Adam:2016vzf, Gudnason:2016kuu} which shows that the quartic term is necessary for black hole solution to exist. Here, we discuss which term gives a BPS submodel for an arbitrary dimensional Skyrme model. The BPS Skyrme submodel itself is a model within the generalized Skyrme model which has the maximum symmetry of the map from spatial space to target space leading to a linear lower bound of energy in the topological degree of the map. It is known that such a lower bound exists in four and three-dimensional spacetime Skyrme model \cite{Faddeev:1976pg, Piette:1994mh, Piette:1994ug}. A five-dimensional BPS Skyrme submodel for a specific spherically symmetric system also has been found in \cite{Fadhilla:2020rig}.  
Our main focus in this work is to find the conditions needed for a generalized Skyrme model in arbitrary spatial dimension \(d\) to satisfy the BPS limit. To do so, we use the extension of the BPS Lagrangian method for the Skyrme model which has been demonstrated for four-dimensional spacetime in \cite{Atmaja:2019gce}. With the conditions for the BPS Skyrmion at hand, we can deduce the BPS bound on the Skyrmion energy and the corresponding first-order differential field equation (Bogomolnyi equation) which could be exploited to derive some features of the BPS Skyrmion. There are two known higher dimensional BPS Skyrme Models, firstly proposed in \cite{Arthur:1996ia} and, in this work, we prove that in spherically symmetric case there are no other submodels that could saturate the BPS limit, this result is demonstrated in section \ref{secBPS}. In the lower spacetime dimensional case the saturation of the BPS limit implies that the total energy is linear to the topological degree (called baryon number), mostly related to the phenomenology of baryonic particles. The linear energy as a function of the baryon number is a well-established fact in nuclear physics \cite{Adam:2010zz}, but for higher-dimensional objects such as \(D\)-brane solitons and strings, such BPS property have a richer implication even though we have not really understood the physical mechanism behind it. We also explore some of these physical properties related to the BPS sector of the Skyrme model as a model for our higher dimensional object.

In this paper we use early alphabets, \(a,b,c,\dots\), as scalar multiplets indices and mid alphabets, \(i,j,k,\dots\) as spacetime indices. Both indices sets take value from \(0\) to \(d\) where \(d+1\) is the spacetime dimensionality. The Einstein summation convention is always to be understood for any repeating indices for both tensors and scalar multiplets. \(\textbf{i}\) represents the imaginary number basis, i.e. \(\textbf{i}=\sqrt{-1}\) and the combinatorial notation \(C^p_q\) reads "\(p\) choose \(q\)" and explicitly given by
\begin{equation}
    C^p_q\equiv\frac{p!}{(p-q)!q!}~,
\end{equation} for \(q\leq p\) and zero for \(q>p\). Because the construction of the theory relies heavily on combinations of some objects such as eigenvalues and winding numbers, it is useful for us to simplify the notations by exploiting elementary symmetric polynomials. The \(k\)-th elementary symmetric polynomials from \(n\) variables are defined as \begin{equation}
    e_k(x_1,x_2,\dots,x_n)\equiv\sum_{i_1\neq i_2\neq \dots \neq i_k}^n \frac{1}{k!} x_{i_1}x_{i_2}\dots x_{i_k}~,
\end{equation}
and we can calculate the expression of \(e_k(x_1,x_2,\dots,x_n)\) for each \(k\) recursively via Girard-Newton formula, namely
\begin{equation}
    k e_k=\sum_{i=1}^k (-1)^{i-1}e_{k-1}p_i
\end{equation}
for all \(n\geq 1\) and \(n\geq k\geq 1\) where we have defined \(p_k\equiv p_k(x_1,x_2,\dots,x_n)\) as
\begin{equation}
    p_k\equiv \sum_{i=1}^n x^k_i~.
\end{equation}
Throughout this paper, we are going to consider the Minkowskian spacetime manifold which has a compact submanifold, given by
\begin{equation}
    ds^2=-dt^2+dr^2+r^2d\Omega_{d-1}^2~.
\end{equation}
The compact submanifold with metric \(d\Omega_{d-1}^2\) resembles a \(S^{d-1}\) which allows \(N=\left\lfloor\frac{d}{2}\right\rfloor\) rotational planes. Explicit choices of the coordinate systems on the manifold of \(d\Omega_{d-1}^2\) are described in a more detailed manner in later sections.
\section{The Skyrme Model}\label{sect:Model}
Let us first consider the standard Skyrme model action functional in \(3+1\) spacetime dimensions given by
\begin{equation}\label{eq:SkyrmeL4d}
    \mathcal{S}_4= \int d^4x\sqrt{-g}\left( \gamma_1g^{ij}~\mathcal{B}_{i}^a \mathcal{B}_j^a+\gamma_2g^{ij}g^{kl}~\mathcal{C}^a_{ik}\mathcal{C}^a_{jl}\right) ~ ,
\end{equation}
with \(\mathcal{B}^a_i=\frac{1}{2\textbf{i}}\text{Tr}\left(U^\dagger\sigma^a\partial_iU\right)\) and \(\mathcal{C}^a_{ij}=\varepsilon^{abc}B^b_iB^c_j\) where \(U=\phi^0I+\textbf{i}(\phi^1\sigma^1+\phi^2\sigma^2+\phi^3\sigma^3)\) is an \(SU(2)\) valued chiral field originally  proposed by T. Skyrme \cite{Skyrme:1961vq, Skyrme:1962vh} and \(\varepsilon^{abc}\) is Levi-Civita symbol.  The vector \(\sigma^a=(\sigma^1,\sigma^2,\sigma^3)\) are Pauli matrices which are the generators of \(SU(2)\), while  $(\phi^a)$ with \(a=0,1,2,3\), are real scalar fields satisfying \(O(4)\) model condition, $ \phi^a\phi^a=1$, see also for example,  \cite{Brihaye:2017wqa, Manton:2004tk, Arthur:1996ia}. 

Now, we introduce the strain tensor which is defined as \(D\equiv JJ^T\) where \(J\) is the Jacobian of generalized harmonic map \(\phi=(\phi^0,\vec{\phi})\). In the recipe given in \cite{manton1987}, we can rewrite the two terms in the Lagrangian of the static case as the invariants of the harmonic map, namely
\(\lambda_1^2+\lambda_2^2+\lambda_3^2\) and \(\lambda_1^2\lambda_2^2+\lambda_1^2\lambda_3^2+\lambda_2^2\lambda_3^2\) with \(\lambda_1^2,\lambda_2^2,\lambda_3^2\) are the eigenvalues of \(D\). We can observe that the use of the strain tensor eigenvalues reduces the problem of finding the BPS bound into an algebraic problem. Thus, we can exploit this method to generalize the static Skyrme model  

Let us discuss the details of this strain tensor for higher dimensional static Skyrme model as follows. For a map from a \(d\) dimensional Euclidean space, \(\mathbb{R}^{d}\), to a \(d\) dimensional sphere, \(S^d\), we have a \(d\times d\) Jacobian matrix \(J\), leading to a \(d\times d\) strain tensor. Suppose we have the eigenvalue of \(D\) from a \(d\) dimensional Skyrme Model as \(\lambda_1^2,\lambda_2^2,\dots,\lambda_d^2\). The corresponding Lagrangian can be written in the most general way as
\begin{equation}
    \mathcal{L}=-\gamma_0 V-\sum_{n=1}^d \gamma_n\mathcal{L}_{2n} ~ ,
\end{equation}
where \(\gamma_i\in\mathbb{R}\) are coupling constants, \(V\equiv V(\phi)\) is an arbitrary potential and \(\mathcal{L}_{2n}\) are combinations of eigenvalues which can be written in elementary symmetric polynomials, namely
\begin{eqnarray}
    \mathcal{L}_{2}&\propto& e_1(\lambda_1^2,\lambda_2^2,\dots,\lambda_d^2) ~ , \nonumber\\ 
    \mathcal{L}_{4}&\propto& e_2(\lambda_1^2,\lambda_2^2,\dots,\lambda_d^2)  ~ , \nonumber\\
    \vdots \nonumber\\
    \mathcal{L}_{2d}&\propto& e_d(\lambda_1^2,\lambda_2^2,\dots,\lambda_d^2) ~ .\label{constLag}
\end{eqnarray}
In static four dimensional spacetime case (\(d=3\)), this gives the generalized Skyrme model which is just the traditional model from Skyrme added with BPS-Skyrme term (\(\mathcal{L}_6\propto \lambda_1^2\lambda_2^2\lambda_3^2\)) proposed in \cite{Adam:2010zz}.\\
By following the prescription given in four dimensional spacetime model we can construct $ \mathcal{L}$ using 
\begin{eqnarray}
    \mathcal{L}_{2n}&\equiv&\frac{1}{(n!)^2}\phi^{a_1}_{[i_1}\dots\phi^{a_n}_{i_n]}\phi^{a_1}_{[j_1}\dots\phi^{a_n}_{j_n]}g^{i_1j_1}\dots g^{i_nj_n} ~ ,
\end{eqnarray}
so that the Lagrangian of the model mentioned above can be written  as
\begin{equation}\label{2}
    \mathcal{L}=-\gamma_0 V-\sum_{n=1}^d \frac{\gamma_n}{\left(n!\right)^2}\phi^{a_1}_{[i_1}\dots\phi^{a_n}_{i_n]}\phi^{a_1}_{[j_1}\dots\phi^{a_n}_{j_n]}g^{i_1j_1}\dots g^{i_nj_n} ~ ,
\end{equation}
where we have defined \(\phi^a_i=\partial_i\phi^a=\frac{\partial\phi^a}{\partial x^i}\) for simplicity.
This is just an \(O(d+1)\) model as the generalization proposed in \cite{Brihaye:2017wqa} if we take the potential to be \(V=1-\phi^0\). As the summary, the action functional for \(d+1\) dimensional Skyrme model takes the form
\begin{equation}\label{SkyrmeAction}
    \mathcal{S}_{d+1}=-\int d^{d+1}x\sqrt{-g}\left[\gamma_0 V+\sum_{n=1}^d \frac{\gamma_n}{\left(n!\right)^2}\phi^{a_1}_{[i_1}\dots\phi^{a_n}_{i_n]}\phi^{a_1}_{[j_1}\dots\phi^{a_n}_{j_n]}g^{i_1j_1}\dots g^{i_nj_n}\right] ~ .
\end{equation}
In fact, the action functional \eqref{SkyrmeAction} does accommodate time dependencies on the fields, but we restrict our model to be spherically symmetric and static. Because the model is static, then \(\phi^a_0=0\) is implied for every possible value of \(a\) which means that the strain tensor effectively lives on the \(d\) dimensional spatial manifold. This fact reduces the representation of strain tensor into a \(d\times d\) matrix, hence, consistent with the construction for \(\phi:\mathbb{R}^d\rightarrow S^d\) we have shown above.
The topological charge density of this higher dimensional spacetime model is given by (up to a normalization constant. See \cite{Arthur:1996ia,Brihaye:2017wqa})
\begin{equation}\label{topDensity}
    \rho\propto\varepsilon^{i_1\dots i_d}\varepsilon^{a_1 \dots a_{d+1}}\phi^{a_1}_{i_1}\dots\phi^{a_d}_{i_d}\phi^{d+1} ~ .
\end{equation}
In the next subsection, we cast the above model into  the combination of the strain tensor \(D\)'s eigenvalue scheme using so called the generalized hedgehog ansatz. The field equations for Skyrmion can be found by substituting the corresponding ansatz to the dynamical equations of the Skyrme model came from the variation of \(\phi^a\) on the Skyrme action functional, namely
            \begin{equation}
                \left(\delta^{cb}-\phi^c\phi^b\right)\left[\nabla_k\left(\sum_{n=1}^d \frac{2\gamma_n}{\left(n!\right)^2}g^{i_1j_1}\dots g^{i_n j_n}\phi^{a_1}_{[i_1}\dots\phi^{a_n}_{i_n]}\frac{\partial}{\partial \phi^b_k}\phi^{a_1}_{[j_1}\dots\phi^{a_n}_{j_n]}\right)-\frac{\partial}{\partial \phi^b} \gamma_0 V\right]=0~.
            \end{equation}
The dynamical field equations above can be simplified by exploiting the Levi-Civita symbol and strain tensor, which lead to the following field equations
\begin{eqnarray}\label{secondFieldEq}
\left(\delta^{cb}-\phi^c\phi^b\right)\left[\sum_{n=1}^d \frac{2n\gamma_n}{n!(d-n)!}\nabla_{j}\left(\phi^{b,i}H_i^j(n,d)\right)-\frac{\partial}{\partial \phi^b} \gamma_0 V\right]=0~,
\end{eqnarray}
Here, we define a new tensor \begin{equation}H_i^j(n,d)\equiv\varepsilon_{i~i_2\dots i_n k_{n+1}\dots k_d}\varepsilon^{j~j_2\dots j_n k_{n+1}\dots k_d}\prod_{m=2}^n D_{j_m}^{i_m}\end{equation} to simplify the computational process, with \(D_{j}^{i}\)s are components of strain tensor \(D\). \(H_i^j(n,d)\) has special features, namely, it is a diagonal tensor if the strain tensor is diagonal. This is the desired property which are going to be exploited more in the calculation of field equations for each submodels. 
\subsection{Spacetime Background and the suitable ansatz}\label{ansatzDetails}
It is necessary to choose in which coordinate the calculations are done because different coordinate requires different ansatz to reduce the complexity of the resulting field equation (without any loss of generality). Here, we consider polyspherical coordinate system\footnote{This coordinate system is an extended version of spherically symmetric Euclidean space that allows more than one independent rotational planes. The details of polyspherical coordinate system's properties and the way we can construct such parametrization for \(\mathbb{R}^d\) are given in a more detailed manner in Appendix \ref{polyspherical}.} with \(N=\left\lfloor\frac{d}{2}\right\rfloor\) azimuthal coordinate \(\varphi_i\)'s, namely
\begin{equation}
    ds^2=-dt^2+dr^2+\sum\limits_{i=1}^Nr^2\left(d\mu_i^2+\mu_i^2d\varphi_i^2\right)
\end{equation}
with \(\sum_{i}\mu_i^2=1\) for even \(d\), and
\begin{equation}
    ds^2=-dt^2+dr^2+r^2d\mu_0^2+\sum\limits_{i=1}^Nr^2\left(d\mu_i^2+\mu_i^2d\varphi_i^2\right)
\end{equation}
with \(\mu_0^2+\sum_{i}\mu_i^2=1\) for odd \(d\). The \(\mu_i\) are called cosine coordinates and we choose to reparameterize them as follows
\begin{eqnarray}
    \mu_0&=&\cos\theta_0~,\nonumber\\
    \mu_1&=&\sin\theta_0\cos\theta_1~,\nonumber\\
    \mu_2&=&\sin\theta_0\sin\theta_1\cos\theta_2~,\nonumber\\
    &&\vdots\nonumber\\
    \mu_{N-1}&=&\sin\theta_0\sin\theta_1\dots\cos\theta_{N-1}~,\nonumber\\
    \mu_{N}&=&\sin\theta_0\sin\theta_1\dots\sin\theta_{N-1}~.
\end{eqnarray}
the even \(d\) case is simply recovered by choosing \(\theta_0=\frac{\pi}{2}\) and it is straightforward to prove that the constraint \(\mu_0^2+\sum_{i}\mu_i^2=1\) is satisfied. 

The spherically symmetric Skyrme field ansatz is a coordinate system on the target space, \(S^d\), given by \begin{equation}
    \phi=(\cos\xi,\Vec{n}\sin\xi)
\end{equation}
with \(\Vec{n}\cdot\Vec{n}=n^an^a=1\) and \(\xi\equiv \xi(r)\) interpolates between \(\pi\) and zero. In this form of ansatz we have the following expression of strain tensor (in matrix form)
\begin{equation}\label{Strain}
    D=\frac{\sin^2\xi}{r^2}M+\left(\xi'{}^2-\frac{\sin^2\xi}{r^2}\right)\frac{\Tilde{\textbf{x}}\Tilde{\textbf{x}}^T}{r^2}
\end{equation}
where components of matrix \(M\) is given by
\begin{equation}
    M_{ij}=\frac{\partial \Tilde{x}^a}{\partial x^i}\frac{\partial \Tilde{x}^a}{\partial x^j}
\end{equation}
with \(x^i\)s are coordinates in the standard Euclidean space, defined in \eqref{EucCoorEveni} for even \(d\) and \eqref{EucCoorOddi} for odd \(d\), and we have defined a new vector \(\Tilde{\textbf{x}}\equiv r \Vec{n}\), hence \(\Tilde{x}^a\) represents component of \(\Tilde{\textbf{x}}\). here, \(i\) runs from \(1\) to \(d\), corresponding to spatial coordinate only.
In order to diagonalize the resulting strain tensor of this map we need to employ the same symmetry features of space-time spherical submanifold's coordinate to the target space coordinate by choosing \(\Vec{n}\) field to be parameterized by the cosine and azimuthal coordinates. By exploiting the normalization constraint of \(\mu_i\)'s we arrive at the following ansatz of \(\Vec{n}\)
\begin{equation}\label{ansatzEven}
    \Vec{n}_{\text{even}}=\left(\mu_1\cos(g_1),~\mu_1\sin(g_1),~\dots,~\mu_N\cos(g_N),~\mu_N\sin(g_N)\right)~.
\end{equation}
for even \(d\) and
 \begin{equation}\label{ansatzOdd}
    \Vec{n}_{\text{odd}}=\left(\mu_0,~\mu_1\cos(g_1),~\mu_1\sin(g_1),~\dots,~\mu_N\cos(g_N),~\mu_N\sin(g_N)\right)~.
\end{equation}
for odd \(d\) where we have defined the functions \(g_i\equiv n_i\varphi_i\) with \(n_i\in\mathbb{Z}\) are the winding numbers of each rotational plane. The constants \(n_i\) is crucial to construct Skyrmion with non-unit topological degree because substitution of \eqref{ansatzEven} and \eqref{ansatzOdd} to \eqref{topDensity} gives the following topological charge
\begin{equation}
    B=\frac{1}{C_d}\int \sqrt{-g}\rho ~d^dx=n_1n_2\dots n_N
\end{equation}
where \(C_d\) is a normalization constant, given that \(B\) must be equal to one when we consider a unit-topological case with all winding numbers equal to one. We can see that the well-known property of Skyrmion topological charge is satisfied where \(B\in\mathbb{Z}\). In fact, this is inevitable because we consider a model which belongs to the homotopy class \(\pi_d(S^d)=\mathbb{Z}\) \cite{Manton:2004tk}. It is important to note that \(n_i\)s themselves are not topological invariants and only their product in the form of topological charge is a topological invariant. Here, \(n_i\)s only describe how we parametrize topological charge in poly spherical coordinate.

We can view the static Skyrme field ansatz as a map \(\mathbb{R}^d\rightarrow S^d\) given by 
\begin{equation}
    (r,\theta_0,\dots,\theta_{N-1},\varphi_1,\dots,\varphi_{N})\rightarrow (\xi,\theta_0,\dots,\theta_{N-1},n_1\varphi_1,\dots,n_N\varphi_{N})~.
\end{equation}
Hence, we can calculate the Jacobian matrix and its corresponding strain tensor explicitly. The resulting eigenvalues of strain tensor within this choice of ansatz are
\begin{equation}\label{EigvalOdd}
    \lambda_{2N+1}^2=\xi'{}^2,~\lambda_{N+1}^2=\dots=\lambda_{2N}^2=\frac{\sin^2\xi}{r^2},\text{ and}~\lambda_{k}^2=n_k^2\frac{\sin^2\xi}{r^2}~\text{for }k=1,2,\dots,N~,
\end{equation}
for odd \(d\) cases and 
\begin{equation}\label{EigvalEven}
    \lambda_{2N}^2=\xi'{}^2,~\lambda_{N+1}^2=\dots=\lambda_{2N-1}^2=\frac{\sin^2\xi}{r^2},\text{ and}~\lambda_{k}^2=n_k^2\frac{\sin^2\xi}{r^2}~\text{for }k=1,2,\dots,N~,
\end{equation}
for even \(d\) cases. This is the expected form of eigenvalues for higher dimensional case which reduce to the one found in \(d=4\) case \cite{Fadhilla:2020rig}. From here, calculating the explicit form of effective Lagrangian by combinations of strain tensor eigenvalues is straightforward via equation \eqref{constLag}. It is found that in five dimensional Skyrme model, there exist an interesting property, known as topological degeneracy, where some multi-soliton states of BPS skyrmion could have different energy states with the same value of topological charge as a result of several possible configurations of winding numbers. This property bring up a question on how winding numbers \(n_i\) fundamentally affect skyrmion properties and the eigenvalues we have for this arbitrary dimensional Skyrme model that depend directly to the winding numbers is a good starting point to do so.

\subsection{The BPS Submodels}\label{secBPS}

From equations \eqref{EigvalOdd}, \eqref{EigvalEven} and \eqref{constLag} we can construct the effective Skyrme Lagrangian of the form
 \begin{equation}
    \mathcal{L}_{eff}=- \left[(\xi')^2\sum_{n=1}^d\gamma_n\frac{\sin^{2(n-1)}\xi}{r^{2(n-1)}}{}_dK_{n-1}+\gamma_0V+\sum_{n=1}^d\frac{\gamma_n\sin^{2n}\xi}{r^{2n}}{}_dK_n\right] ~,
\end{equation}
where we have defined functions of winding number \({}_dK_n\equiv{}_dK_n(n_1,\dots n_{N})~,\) which also depend on spatial dimensions \(d\), given by
\begin{itemize}
    \item For odd \(d\) cases
    \begin{equation}
        {}_dK_n=\sum_{i=0}^{n}C^{(d-1)/2}_{n-i}e_i(n_1^2,n_2^2,\dots,n_{(d-1)/2}^2)=\sum_{i=0}^{n}C^{N}_{n-i}e_i(n_1^2,n_2^2,\dots,n_N^2)~.
    \end{equation}
    \item For even \(d\) cases
    \begin{equation}
        {}_dK_n=\sum_{i=0}^{n}C^{(d/2)-1}_{n-i}e_i(n_1^2,n_2^2,\dots,n_{d/2}^2)=\sum_{i=0}^{n}C^{N-1}_{n-i}e_i(n_1^2,n_2^2,\dots,n_N^2)~.
    \end{equation}
\end{itemize}
The construction of \({}_dK_n\) from eigenvalues of strain tensor is straightforward but could be very cumbersome for higher values of \(d\), hence we carry the calculations only for some special cases where their explicit form is necessary.
These functions satisfies
\begin{equation}
    {}_dK_n\geq C^{d}_{n}-C^{d-1}_{n-1}~.
\end{equation}
with equality is achieved only if all winding numbers are equal to one, \(n_1=\dots=n_{N}=1\). To find the BPS type solution of this model we can use BPS Lagrangian method for Skyrme model which can be found on \cite{Atmaja:2019gce,Stepien:2018mti}. Since our metric satisfies \(\sqrt{-g}\propto r^{d-1}\), the method suggests a BPS Lagrangian \(\mathcal{L}_{BPS}=-\frac{Q}{r^{d-1}}\xi'\) with \(Q\equiv Q(\xi)\) does not depend explicitly on the coordinates. At the BPS limit \(\mathcal{L}_{eff}-\mathcal{L}_{BPS}=0\) we have 
\begin{equation}\label{15}
    \left[\sum_{n=1}^d\gamma_n\frac{\sin^{2(n-1)}\xi}{r^{2(n-1)}}{}_dK_{n-1}\right](\xi')^2-\left[\frac{Q}{r^{d-1}}\right]\xi'+\left[\gamma_0V+\sum_{n=1}^d\frac{\gamma_n\sin^{2n}\xi}{r^{2n}}{}_dK_n\right]=0 ~ ,
\end{equation}
which is the quadratic equation of \(\xi'\). Solving \eqref{15} for \(\xi'\) and then imposing condition for which we have only single solution, it gives
\begin{equation}\label{16}
    \frac{Q^2}{4r^{2(d-1)}}-\left[\sum_{n=1}^d\gamma_n\frac{\sin^{2(n-1)}\xi}{r^{2(n-1)}}{}_dK_{n-1}\right]\left[\gamma_0V+\sum_{n=1}^d\frac{\gamma_n\sin^{2n}\xi}{r^{2n}}{}_dK_{n}\right]=0 ~ .
\end{equation}
Equation \eqref{16} is nothing but a polynomials in \(r\), thus, by setting every corresponding coefficient to be zero, we have a set of equations to determine each \(\gamma_n\) that satisfies the BPS limit. These equations are
\begin{eqnarray}
0={}_dK_{k-1}\gamma_0\gamma_kV+\sum_{n+m=k}\gamma_n\gamma_m~{}_dK_{n-1}~{}_dK_m ~~~&&\text{For }k=2,3,\dots,d~ ,\\
0=\sum_{n+m=k}\gamma_n\gamma_m~{}_dK_{n-1}~{}_dK_m ~~~&&\text{For }k=d+1,d+2,\dots,2d~ ,\\
\sin^{2(d-1)}\xi\left[\gamma_0\gamma_dV+\sum_{n+m=d}\gamma_n\gamma_m~{}_dK_{n-1}~{}_dK_m\right] &=&\frac{Q^2}{4}~ .
\end{eqnarray}
Since \({}_dK_n\) are positive definite quantities, we can deduce that for all \(d\in\mathbb{N}\) there exists only one BPS submodel with \(\gamma_1=\dots=\gamma_{d-1}=0\). This submodel is a higher dimensional extension of BPS-Skyrme model in four spacetime dimensions \cite{Adam:2010zz}. On the other side, there is another BPS submodel if we restrict our case for only even \(d\) with coupling constants \(\gamma_0=\dots=\gamma_{\frac{d}{2}-1}=\gamma_{\frac{d}{2}+1}=\dots=\gamma_d=0\), i.e. only \(\gamma_{\frac{d}{2}}\) is non-zero. This submodel is also a higher dimensional extension of the model in five spacetime dimensions introduced in \cite{Brihaye:2017wqa}. Both submodels in \(d+1\) dimensional spacetime  are constructed from the scale invariant hierarchy and the scalar norm of the topological current (or topological charge density). Solutions and properties of both BPS submodels are discussed in the next sections.

\section{BPS Bounds and BPS Type Solutions Via Bogomolnyi Equations}
\subsection{BPS Bounds of The First and Second Submodel}
Firstly, let us consider the first submodel that exist for any value of spatial dimension \(d\in\mathbb{N}\). In this case we only have one BPS submodels as mentioned above. The Lagrangian of this submodel is given by
\begin{equation}\label{BPS1}
    \mathcal{L}_I= -\gamma_0 V - \frac{\gamma_d}{\left(d!\right)^2}\phi^{a_1}_{[i_1}\dots\phi^{a_d}_{i_d]}\phi^{a_1}_{[j_1}\dots\phi^{a_d}_{j_d]}g^{i_1j_1}\dots g^{i_dj_d} ~ ,
\end{equation}
which gives the following equation of motions for the fields
\begin{equation}
    \left(\delta^{cb}-\phi^c\phi^b\right)\left[\nabla_k\left( \frac{2\gamma_d}{\left(d!\right)^2}g^{i_1j_1}\dots g^{i_d j_d}\phi^{a_1}_{[i_1}\dots\phi^{a_d}_{i_d]}\frac{\partial}{\partial \phi^b_k}\phi^{a_1}_{[j_1}\dots\phi^{a_d}_{j_d]}\right)-\frac{\partial}{\partial \phi^b} \gamma_0 V\right]=0 ~ .
\end{equation}
The BPS inequality can be found by introducing the energy functional of the static case which satisfies \(E= \int\sqrt{-g}dx^d\mathcal{L}\). Recall that the second term in \eqref{BPS1} is just the determinant of the strain tensor, thus our energy functional takes the following form
\begin{equation}
    E=\int\sqrt{-g}d^dx\left[\gamma_d\text{det}D+\gamma_0V\right]=\int\sqrt{-g}d^dx\left[\left(\sqrt{\gamma_d\text{det}D}\mp\sqrt{\gamma_0V}\right)^2\pm2\sqrt{\gamma_0\gamma_dV\text{det}D}\right] ~ .
\end{equation}
In the second term we can transform the spacetime invariant measure \(d^dx\sqrt{-g}\) to the target space invariant measure by using \(\sqrt{\text{det}D}\sqrt{-g}~d^dx=\sqrt{-g}\left|\det{J}\right|d^dx=\sqrt{h}~d^d\eta\) where \(\eta^a\)'s are the coordinates and \(h\) is the metric tensor determinant of the target manifold. The integration on the entire target manifold is actually \(B\) fold of integration on the entire \(S^d\), where \(B\) is the topological degree of \(\phi\), hence we state the resulting BPS inequality
\begin{equation}\label{BPSbound1}
    E\geq 2B\sqrt{\gamma_0\gamma_d}\text{Vol}S^d\left<\sqrt{V}\right>_{S^d} ~ ,
\end{equation}
with \(\text{Vol}S^d\) is the total volume of compact manifold \(S^d\) and \(\left<a\right>_{S^d}\) denotes the average of function \(a\) on \(S^d\). The straightforward transformation of base space invariant measure (Volume form) to the target space measure implies that the derivative term of this submodel is invariant under all volume-preserving diffeomorphisms on \(S^d\), which has been shown in \cite{Adam:2008uj}. Because this model belong to the same class of BPS Skyrme model in \cite{Adam:2012sa} and generally (for higher dimensions) in \cite{Adam:2008uj}, then we can conclude that the integrability is guaranteed. The nonzero potential term itself provides a symmetry braking into a smaller group which leaves the \(\xi\) unchanged, represented by the constraint \(\Vec{n}\cdot\Vec{n}=1\). This restricts the symmetries further for only volume-preserving diffeomorphisms on \(S^{d-1}\), which again, implies that this model is a straightforward extension of the model in \cite{Adam:2010fg}. 

The BPS bound which restricts the lower bound of energy to be linear in topological degree is the main feature of this model. This BPS bound given in \eqref{BPSbound1} strongly constrains the energy spectrum of this type of soliton which belongs to the class of \(D\)-brane soliton. The bound is satisfied by generalized Skyrme as well since adding the other terms only makes the energy higher. Furthermore, the result \eqref{BPSbound1} implies that for every \(d\in\mathbb{N}\) there is at least one class of Skyrmion (or a submodel) that is non-interacting, provided that the BPS bound is saturated. This fact not only applies to the compacton solution but also the regular ones, as we show in a later subsection that regular solutions exist and are well-behaved.

Now, let us consider the special case of even \(d\). For this case there exists two type of BPS submodels. The first one is from the general \(d\in\mathbb{N}\) case which has been previously described above and the second one is obtained by taking only \(\gamma_{\frac{d}{2}}\) to be non-zero. The Lagrangian of the second submodel is given by
\begin{equation}\label{BPS2}
    \mathcal{L}_{II}=-  \frac{\gamma_{d/2}}{\left((d/2)!\right)^2}\phi^{a_1}_{[i_1}\dots\phi^{a_{d/2}}_{i_{d/2}]}\phi^{a_1}_{[j_1}\dots\phi^{a_{d/2}}_{j_{d/2}]}g^{i_1j_1}\dots g^{i_{d/2}j_{d/2}}
\end{equation}
that implies the following field equations (Euler-Lagrange equations)
\begin{equation}\label{fieldEqSub2}
    \left(\delta^{cb}-\phi^c\phi^b\right)\nabla_k\left(g^{i_1j_1}\dots g^{i_{d/2} j_{d/2}}\phi^{a_1}_{[i_1}\dots\phi^{a_{d/2}}_{i_{d/2}]}\frac{\partial}{\partial \phi^b_k}\phi^{a_1}_{[j_1}\dots\phi^{a_{d/2}}_{j_{d/2}]}\right)=0
\end{equation}
Here, we take a slightly different approach to find the BPS inequality of this submodel. We know that from the construction of Lagrangian terms via strain tensor eigenvalues we have
\begin{equation}\label{Leven}
    \mathcal{L}_d\propto e_{d/2}(\lambda_1^2,\dots,\lambda_d^2)\propto\lambda_1^2\dots\lambda_{\frac{d}{2}}^2+\dots+\lambda_{\frac{d}{2}+1}^2\dots\lambda_d^2
\end{equation}
There are \(C^d_{d/2}\) terms in \eqref{Leven}, hence by employing arithmetic-geometric means inequality, which reads \(\sum_i w_i f_i\geq\prod_i f_i^{w_i}\) with \(\sum_i w_i=1\), to \eqref{Leven} with identical \(w_i=\frac{1}{C^d_{d/2}}\) for every term we can proceed as follow
\begin{eqnarray}
&&\lambda_1^2\dots\lambda_{\frac{d}{2}}^2+\dots+\lambda_{\frac{d}{2}+1}^2\dots\lambda_d^2\nonumber\\
&=&C^d_{d/2}\left(\frac{1}{C^d_{d/2}}\lambda_1^2\dots\lambda_{\frac{d}{2}}^2+\dots+\frac{1}{C^d_{d/2}}\lambda_{\frac{d}{2}+1}^2\dots\lambda_d^2\right)\nonumber\\
&\geq&C^d_{d/2}(\lambda_1^2)^{\frac{C^{d-1}_{(d/2)-1}}{C^d_{d/2}}}\dots(\lambda_d^2)^{\frac{C^{d-1}_{(d/2)-1}}{C^d_{d/2}}}\nonumber\\
&=&C^d_{d/2}\lambda_1\dots\lambda_d=C^d_{d/2}\sqrt{\text{det}D}
\end{eqnarray}
From here, it is straightforward to conclude that this BPS submodel satisfies a BPS inequality of the form
\begin{equation}\label{BPSbound2}
    E\geq B~\gamma_{d/2}C^d_{d/2}\text{vol}S^d
\end{equation}
with BPS limit saturation happens if and only if \(\lambda_1^2=\dots=\lambda_d^2\), i.e. all the eigenvalues are identical. This is the self-duality condition in terms of strain tensor eigenvalues. In fact, this property can be derived directly from the usual self-dual equation for \(O(d+1)\) Skyrme-sigma (see, \cite{Arthur:1996ia}) model given by
\begin{equation}\label{self-dualSubmodel2}
    \phi^{a_1}_{[i_1}\dots\phi^{a_{d/2}}_{i_{d/2}]}=\frac{1}{((d/2)!)^2}\varepsilon_{i_1\dots i_d}\varepsilon^{a_1\dots a_{d+1}}\phi^{a_{(d/2)+1}}_{[j_{(d/2)+1}}\dots\phi^{a_{d}}_{j_{d}]}\phi^{a_{d+1}} g^{i_{(d/2)+1}j_{(d/2)+1}}\dots g^{i_{d}j_{d}}.
\end{equation}
that reduces into \(\frac{C^{d}_{(d/2)}}{2}\) equations which can be expressed as follows
\begin{equation}\label{self-dualSubmodel2Eig}
    \lambda_{k_1}\dots\lambda_{k_{d/2}}\pm\lambda_{k_{(d/2)+1}}\dots\lambda_{k_{d}}=0,
\end{equation}
with all the indices \(k_i\) must take different values, \(k_i\in{1,2,\dots,d}\).  We can show that equation \eqref{self-dualSubmodel2Eig} is equivalent to \begin{equation}
    \lambda_1^2=\dots=\lambda_d^2,
\end{equation}
as follows. Let us choose two arbitrary \(\lambda_{k_i}\) from the left-hand-side and \(\lambda_{k_j}\) from the right-hand-side of \(\lambda_{k_1}\dots\lambda_{k_i}\dots\lambda_{k_{d/2}}=\mp\lambda_{k_{(d/2)+1}}\dots\lambda_{k_j}\dots\lambda_{k_{d}}\). By switching \(\lambda_{k_i}\) to the right-hand-side and \(\lambda_{k_j}\) to the left-hand-side we have \(\lambda_{k_1}\dots\lambda_{k_j}\dots\lambda_{k_{d/2}}=\mp\lambda_{k_{(d/2)+1}}\dots\lambda_{k_i}\dots\lambda_{k_{d}}\), and since all \(k_i\)s are different, this equation is also allowed by \eqref{self-dualSubmodel2Eig}. Dividing the first equation with the switched equation gives
\begin{equation}
\frac{\lambda_{k_i}}{\lambda_{k_j}}=\frac{\lambda_{k_j}}{\lambda_{k_i}},
\end{equation}
that implies \(\lambda_{k_i}^2=\lambda_{k_j}^2\). Since \(k_i\) and \(k_j\) are arbitrary, the equation applies to all \(k_i\in{1,2,\dots,d}\). This shows the equivalence between self-dual equation and BPS limit saturation conditions in terms of strain tensor eigenvalues.

Lastly, it is beneficial to show that \eqref{BPSbound2} is indeed the true lower bound of the energy. To do this, consider the following squared quantity from the self-duality equation
\begin{equation}
    \left(\lambda_{k_1}\dots\lambda_{k_{d/2}}\pm\lambda_{k_{(d/2)+1}}\dots\lambda_{k_{d}}\right)^2=\left(\lambda_{k_1}\dots\lambda_{k_{d/2}}\right)^2+\left(\lambda_{k_{(d/2)+1}}\dots\lambda_{k_{d}}\right)^2\pm2\lambda_1\lambda_2\dots\lambda_d,
\end{equation}
where we have used the fact that all \(k_i\)s must take different values that leads to the last term in the right-hand side become two times the product of all
\(\lambda_i\)s. Both the \(\pm\) sign in the left-hand and right-hand sides take the same value simultaneously with positive sign represents anti-soliton (with negative topological degree), and vice versa. Thus, we proceed with negative sign for the above equation since we are only dealing with soliton (with positive topological degree). Next, summing all of the above \(\frac{C^d_{d/2}}{2}\)  possible terms gives
\begin{equation}
    \sum_{k_1\neq k_2\neq \dots k_d}\left(\lambda_{k_1}\dots\lambda_{k_{d/2}}-\lambda_{k_{(d/2)+1}}\dots\lambda_{k_{d}}\right)^2=e_{d/2}(\lambda_1^2,\dots,\lambda_d^2)-C^2_{d/2}\lambda_1\lambda_2\dots\lambda_d.
\end{equation}
Notice that the first term in the right-hand side is proportional to \(\mathcal{L}_d\), thus we can write the total energy as
\begin{equation}
    E=\gamma_{d/2}\int\sqrt{-g}\left[\sum_{k_1\neq k_2\neq \dots k_d}\left(\lambda_{k_1}\dots\lambda_{k_{d/2}}-\lambda_{k_{(d/2)+1}}\dots\lambda_{k_{d}}\right)^2+C^2_{d/2}\lambda_1\lambda_2\dots\lambda_d\right]~d^dx.
\end{equation}
We can see that if we impose the self-dual equation to the above equation then the summation vanishes, which leave us only the lower-bound term that have been previously found via arithmetic-geometric inequality.

\subsection{BPS Solutions}
\subsubsection{The First Submodel}
The first BPS submodel within the ansatz of this work for the static case (\(\omega=0\)) is given by
\begin{equation}
    E= \int\sqrt{-g}\left[\gamma_dB^2\frac{\sin^{2(d-1)}\xi}{r^{2(d-1)}}\xi'^2+\gamma_0V\right]d^dx~
\end{equation}
where we have substituted the value of topological charge, \(B=n_1n_2\dots n_N\). The BPS limit is satisfied when \(\sqrt{\gamma_d\text{det}D}\mp\sqrt{\gamma_0V}=0\) which leads to
the Bogomolny equation of this submodel, given by
\begin{equation}\label{Bog1B}
    B\frac{\sin^{d-1}\xi}{r^{d-1}}\xi'\mp\sqrt{\frac{\gamma_0}{\gamma_d}}\sqrt{V}=0 ~ .
\end{equation}
which implies that the cut of radius now scales as \(r_{cut-off}\propto B^{\frac{1}{d}}\).\footnote{Similar to the property of BPS Skyrme model in \(d=3\) spatial dimensions} This property of cut-off radius is expected from the scaling property of the model itself. This can be physically interpreted as the volume of the first type of BPS Skyrmions is proportional to its topological degree. 

\begin{figure}[h]
    \centering
    \includegraphics[width=0.8\textwidth]{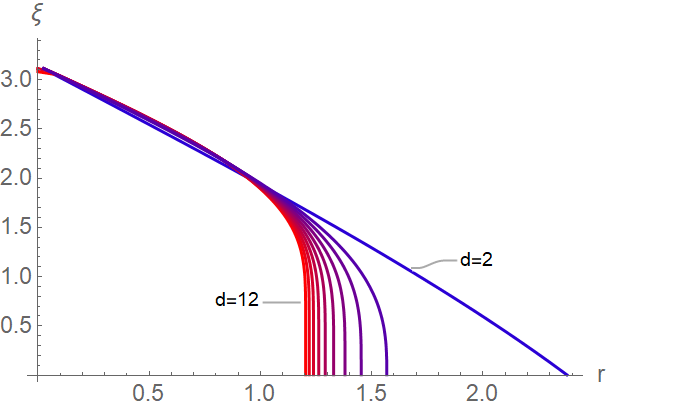}
    \caption{Numerical solution for equation \eqref{Bog1} with pion-mass type potential}
    \label{fig:BPS-Skyrme-D-Dim}
\end{figure}

\begin{figure}[h]
    \centering
    \includegraphics[width=0.8\textwidth]{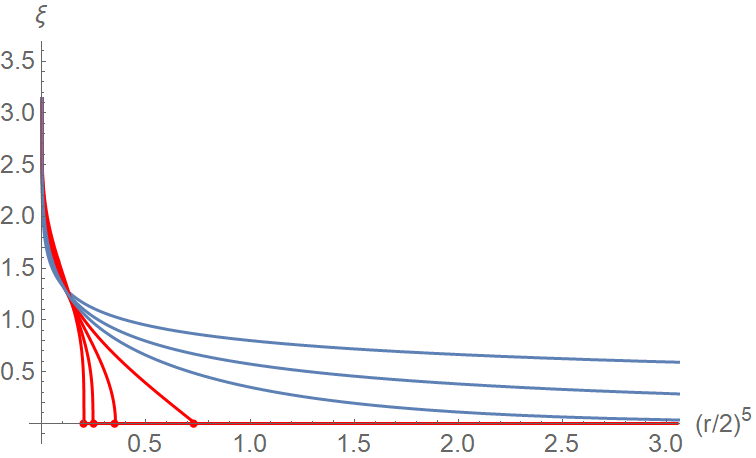}
    \caption{Numerical solutions for equation \eqref{Bog1} with pion-mass type potential to the power \(s=1,2, \dots,8\) for \(d=5\) spatial manifold. The red curves represent \(s=1,2,3,4\) cases, and the blue curves represent \(s=5,6,7\) cases. We can see that for higher \(s\) the profile of \(\xi\) decays slower as \(r\rightarrow\infty\)}
    \label{fig:BPS-Skyrme-D-Dim transition N=d}
\end{figure}

\begin{figure}[h]
    \centering
    \includegraphics[width=0.8\textwidth]{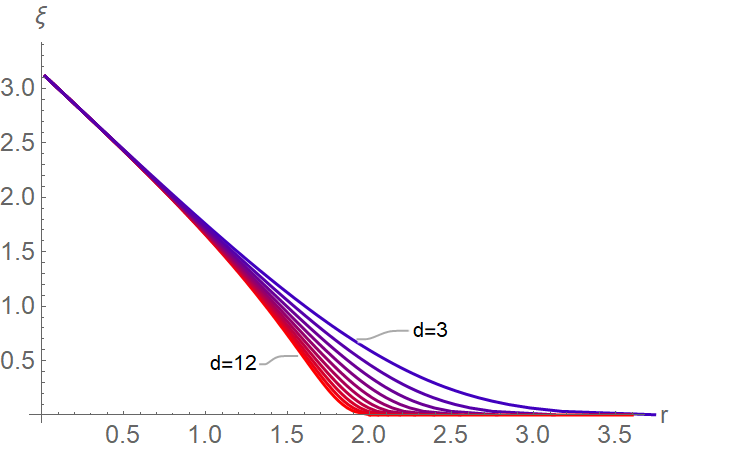}
    \caption{Numerical solutions for equation \eqref{Bog1} with pion-mass type potential to the power of \(s=d\).}
    \label{fig:BPS-Skyrme-D-Dim N=d}
\end{figure}

The Bogomolnyi equation for this model gives an identical solution with the \(B=1\) solution in \eqref{18} and \eqref{18'} if we rescale the radial coordinate as \(r\rightarrow r/B^{\frac{1}{d}}\), hence, there is no new qualitative feature of this non-unit topological charge case. Thus, from this point on, we restrict ourselves to the case of \(B=1\) without any loss of generality which satisfies the following Bogomolnyi equation
\begin{equation}\label{Bog1}
    \frac{\sin^{d-1}\xi}{r^{d-1}}\xi'\mp\sqrt{\frac{\gamma_0}{\gamma_d}}\sqrt{V}=0 ~ .
\end{equation}
We can immediately see that vacuum solution is the trivial solution of this submodel. The non-trivial solution needs a specific form of potential, in this work we use the pion-mass type potential to the power of \(s\) which takes the form \(V=(1-\phi_0)^s=(1-\cos\xi)^s\) with \(s\in\mathbb{R}^+\). Integrating equation \eqref{Bog1} with mentioned pion-mass type potential gives
\begin{equation}\label{18}
    \frac{\sin^{d-s}\left(\frac{\xi}{2}\right)}{d-s}~{}_{2}\text{F}_{1}\left(1-\frac{d}{2},~\frac{d-s}{2};~\frac{2+d-s}{2};~\sin^2\left(\frac{\xi}{2}\right)\right)\mp\sqrt{\frac{2^s\gamma_0}{d^2\gamma_d}}\left(\frac{r}{2}\right)^d+c_d=0 ~ ,
\end{equation}
for \(d\neq s\) or
\begin{equation}\label{18'}
    \frac{\cos^{d}\left(\frac{\xi}{2}\right)}{d}~{}_{2}\text{F}_{1}\left(1,~\frac{d}{2};~\frac{2+d}{2};~\cos^2\left(\frac{\xi}{2}\right)\right)\mp\sqrt{\frac{2^d\gamma_0}{d^2\gamma_d}}\left(\frac{r}{2}\right)^d+c'_d=0 ~ ,
\end{equation}
for special case where \(d=s\) with \({}_{2}\text{F}_{1}\) is the hypergeometric function while \(c_d\) and \(c'_d\) are integration constants. Equation \eqref{18} holds for all \(d\in\mathbb{N}\) because the hypergeometric series from \({}_{2}\text{F}_{1}\) always terminate at finite term. 
From equation \eqref{18} we can find the cut -off value of radius \(r\) in which \(\xi\) takes vacuum value. This cut-off radius is
\begin{equation}
    r_{cut-off}=\begin{cases}2\left[\frac{d}{d-s}\sqrt{\frac{\gamma_d}{2^s\gamma_0}}\frac{\Gamma\left(\frac{d}{2}\right)\Gamma\left(\frac{2+d-s}{2}\right)}{\Gamma\left(d-\frac{s}{2}\right)}\right]^{\frac{1}{d}}& \text{for}~ 0<s<d\\\infty & \text{for}~ s\geq d\end{cases} ~ ,
\end{equation}
which implies that Skyrmion distributed in a smaller radius for higher dimensions and the solution becomes a regular solution for \(s\geq d\).  It is also can be observed that the cut-off radius is asymptotic to 1 as \(d\) goes to infinity for compacton solutions. This property ensures the existence of positive finite energy of compacton solutions in BPS Skyrme model for any dimensions.
Since equations \eqref{18} and \eqref{18'} are transcendental in \(\xi\) we use numerical method to find the solutions for some special cases. The first case is the ordinary pion-mass potential where \(s=1\). This case is a higher dimensional extension of the four-dimensional case in \cite{Adam:2010zz}. We can see in Figure \ref{fig:BPS-Skyrme-D-Dim} that all solutions are of compacton type for \(d\geq2\), in agreement with the special four-dimensional spacetime case.
 The next special case is \(s=d\) where the solution starts to become a regular solution that extends to infinity. We demonstrate this feature in Figure \ref{fig:BPS-Skyrme-D-Dim transition N=d} where we plot the solutions in \(d=5\) spatial manifold for \(s=1,2, \dots,8\). In Figure \ref{fig:BPS-Skyrme-D-Dim N=d} we show the solution profiles for some value of \(d\). We can observe that the solutions become closer as \(d\) goes to infinity which is the same feature we have found in the compacton solutions above. Furthermore, there are no intersections between solutions with different value of \(d\) and all solutions are monotonically decreasing, i.e. \(\xi_d\geq \xi_{d+1}\geq 0\) for \(r\in[0,\infty)\) with \(\xi_d\) is the solution of \(\xi(r)\) in \(d\) spatial dimension. From these mentioned features and further support by the fact that energy for \(d=1\) case is finite, we can conclude that the energy, in general, is positive and finite for all values of \(d\).
Now, consider the energy-momentum tensor of this submodel which is given by
\begin{eqnarray}\label{EnMomTen1}
    T_{ij}&=&2 \frac{d\gamma_d}{\left(d!\right)^2}\phi^{a_1}_{[i_1}\dots\phi^{a_{d-1}}_{i_{d-1}}\phi^{a_d}_{i]}\phi^{a_1}_{[j_1}\dots\phi^{a_{d-1}}_{j_{d-1}}\phi^{a_d}_{j]}g^{i_1j_1}\dots g^{i_{d-1}j_{d-1}}\nonumber\\
    &&-g_{ij}\gamma_0 V-g_{ij} \frac{\gamma_d}{\left(d!\right)^2}\phi^{a_1}_{[i_1}\dots\phi^{a_d}_{i_d]}\phi^{a_1}_{[j_1}\dots\phi^{a_d}_{j_d]}g^{i_1j_1}\dots g^{i_dj_d}
\end{eqnarray}
Substituting the Bogomolnyi equation to \eqref{EnMomTen1} gives
\begin{equation}
    T_{00}=\varepsilon,~T_{0j}=T_{ij}=0~\text{for all values of}~i,j~\text{except 0,}
\end{equation}
where \(\varepsilon\) is the energy density. Here, we can observe that the energy density does not have a simple expression in \(r\), but the asymptotic expression is useful in energy estimate. Suppose that we choose a large positive real \(R\) such that for \(r>R\) the energy density can be approximated by its asymptotic form, namely
\begin{equation}
    \varepsilon\approx \frac{2\gamma_0}{2^s}\xi^{2s}\approx\frac{2\gamma_0}{2^s}\left(2^{s/2}\sqrt{\frac{\gamma_d}{\gamma_0}}\frac{d}{s-d}\right)^{\frac{2s}{s-d}}\frac{1}{r^{\frac{2sd}{s-d}}}~.
\end{equation}
Note that we have used the asymptotic expression of \eqref{Bog1} to find the asymptotic solution of \(\xi\) and we only consider the case with \(s>d\) where asymptotic vacuum is slower to be reached for higher \(s\).\footnote{For the cases with \(s\in[1,d]\) the profiles decay faster as \(r\rightarrow\infty\), thus they have finite total energy if the cases with higher \(s\) have finite total energy.} If \(R\) is arbitrary, then the total energy is always smaller than the following integral
\begin{eqnarray}
\int_0^R\sqrt{-g}\varepsilon~d^dx+\frac{2\gamma_0}{2^s}\left(2^{s/2}\sqrt{\frac{\gamma_d}{\gamma_0}}\frac{d}{s-d}\right)^{\frac{2s}{s-d}}~\text{Vol}S^{d-1}\int_R^\infty r^{d-1}\left(\frac{1}{r^{\frac{2sd}{s-d}}}\right) dr~.
\end{eqnarray}
To have finite energy the second term above must be finite which implies that \( d-1-\frac{2sd}{s-d}<-1 \) or, in other words, the total energy is finite for every \(s\in(d,\infty)\). In a conclusion, the BPS submodel with unit topological degree given in \eqref{BPS1} is equipped with the pion-mass potential to the power \(s\) and static Skyrme field ansatz has finite total energy which is bounded from below by \eqref{BPSbound1}.
\subsubsection{The Second Submodel}
The second submodel has the following energy functional
\begin{equation}
    E=\gamma_{d/2}\int\sqrt{-g}\left[{}_dK_{(d/2)-1}\xi'^2+{}_dK_{d/2}\frac{\sin^{2}\xi}{r^{2}}\right]\frac{\sin^{d-2}\xi}{r^{d-2}}~d^dx~.
\end{equation}
In general, the Bogomolnyi equation of this submodel is non-self-dual which takes the form
\begin{equation}\label{Bog2Eq}
    \xi'\pm \sqrt{\frac{{}_dK_{d/2}}{{}_dK_{(d/2)-1}}}\frac{\sin\xi}{r}=0.
\end{equation}
We can see that this equation is, indeed, non self-dual since the self-duality equation \eqref{self-dualSubmodel2} is not satisfied for \(d>2\) unless we take \(B=1\), but the solution of \eqref{Bog2Eq} is the skyrmion solution of second submodel which satisfies the field equations as demonstrated in appendix \ref{AppB2}. Thus, the resulting solutions are higher topological degree skyrmions that admit spherical symmetry. 

The Bogomolnyi equation of this second submodel satisfies \eqref{fieldEqSub2} and can be derived directly from effective lagrangian via BPS Lagrangian method. The solution of \eqref{Bog2Eq} is given by
\begin{equation}
    \xi=2\arctan\left(\left(\frac{r_0}{r}\right)^{\sqrt{\frac{{}_dK_{d/2}}{{}_dK_{(d/2)-1}}}}\right)~.
\end{equation}
This solution admits topological degeneracy where different solutions and energies exist for a single value of the topological degree.\footnote{We take the positive signed solution of \eqref{Bog2Eq} in order to place the potential vacuum at spatial infinity, resulting a Skyrmion with positive topological degree. In fact, the negative signed solution also exist which is known as anti-Skyrmion with quite similar properties with ordinary Skyrmion} This is because we can choose different configurations of \(n_1\) to \(n_N\) without changing the value of the topological degree. In contrast with the topological degree which is just a product of all winding numbers, \(n_1\) to \(n_N\), the solution and energy highly depend on the value of \({}_dK_n\) which gives different values for different configurations of winding numbers \cite{Fadhilla:2020rig}.

The energy momentum tensor of the second submodel is \begin{eqnarray}\label{EnMomTen2}
    T_{ij}&=& \frac{d\gamma_{d/2}}{\left((d/2)!\right)^2}\phi^{a_1}_{[i_1}\dots\phi^{a_{d/2}}_{i]}\phi^{a_1}_{[j_1}\dots\phi^{a_{d/2}}_{j]}g^{i_1j_1}\dots g^{i_{(d/2)-1}j_{(d/2)-1}}\nonumber\\
    &&-g_{ij} \frac{\gamma_{d/2}}{\left((d/2)!\right)^2}\phi^{a_1}_{[i_1}\dots\phi^{a_{d/2}}_{i_{d/2}]}\phi^{a_1}_{[j_1}\dots\phi^{a_{d/2}}_{j_{d/2}]}g^{i_1j_1}\dots g^{i_{d/2}j_{d/2}}~.
\end{eqnarray}
If we substitute the first order field equation \eqref{Bog2Eq} the trace of energy-momentum tensor of this submodel becomes
\begin{equation}\label{EnergDenSUb2}
    T_{ij}g^{ij}=T_0^0=-\varepsilon=-2^{d+1}\gamma_{d/2}~{}_dK_{d/2}~\left(\frac{r^{\alpha-1}r_0^{\alpha}}{r_0^{2\alpha}+r^{2\alpha}}\right)^d~,
\end{equation}
because the radial component of energy-momentum tensor vanishes, \(T_{1}^1=0\) and the angular component, \(T_2^2,\dots,T_d^d\), are non-zero but the summation of all angular component vanishes, i.e. \(\sum_{i=2}^d T^i_i=0\). In \eqref{EnergDenSUb2} we have defined \(\alpha\equiv\sqrt{\frac{{}_dK_{d/2}}{{}_dK_{(d/2)-1}}}\) to simplify the expression. Now, consider the expression of total energy of this second submodel which is given by
\begin{equation}
    E=2\gamma_{d/2}\sqrt{{}_dK_{(d/2)-1}~{}_dK_{d/2}}\text{Vol}S^d.
\end{equation}
We can see that this energy is higher than the BPS bound due to the non-self-duality of the solution. As a result, there are possible repulsive interactions between \(B=1\) Skyrmions in a multi-soliton configuration where the total topological degree is more than one, which is a similar property found in \(d=4\) case \cite{Fadhilla:2020rig}. 

Interestingly, for \(d=2\) the Bogomolnyi equation \eqref{Bog2Eq} becomes self-dual because it satisfies \begin{equation}
    \lambda_1^2=\xi'^2=\lambda_2^2=n_1^2\frac{\sin^2\xi}{r^2}=\frac{{}_2K_{1}(n_1)}{{}_2K_{0}(n_1)}\frac{\sin^2\xi}{r^2}~.
\end{equation}
This lower dimensional situation cannot be found in a higher-dimensional case, at least, within the same construction as has been shown here. Hence, \(d=2\) is the only case where we can find stable non-interacting BPS Skyrmions of the second BPS submodel for the case of \(B>1\). The stability of this higher topological degree case is further supported by the fact that \(d=2\) is the only case where all components of the energy-momentum tensor vanishes except the \(0,0\) component, \(T^0_0\). This came from the fact that \(d=2\) dimensional space has only one azimuthal coordinate and no polar coordinate, which leads to only one winding number. The second submodel of \(d=2\) we have discussed here is well-known as \(O(3)\) sigma model in \(2+1\) dimensional spacetime and it has been proven in the previous section that this sigma model is a BPS submodel of Baby-Skyrme model, proposed in \cite{Piette:1994mh, Piette:1994ug}. 

Now, let us consider the case where Bogomolnyi equation coincides with self-duality equation.\footnote{The self-dual conditions in terms of strain tensor eigenvalues is given by \(\lambda_1^2=\lambda_2^2=\dots=\lambda_d^2\). Substituting the ansatz from \eqref{ansatzDetails} to the self-dual equation for \(d>2\) gives \(\xi'{}^2=\frac{\sin^2\xi}{r^2}=n_1^2\frac{\sin^2\xi}{r^2}=\dots=n_N^2\frac{\sin^2\xi}{r^2}\) that can only be satisfied for \(B=1\).} In this submodel, this can only be done by taking
\begin{equation}
    \lambda_1^2=\lambda_2^2=\dots=\lambda_d^2~,
\end{equation}
which lead to a \(B=1\) BPS Skyrmion unless \(d=2\). The Corresponding Bogomolny equation becomes
\begin{equation}\label{Bog2}
    \xi'\pm\frac{\sin\xi}{r}=0~.
\end{equation}
Features and solutions of this submodel are well-known and has been discussed in detail in \cite{Brihaye:2017wqa,Fadhilla:2020rig}, namely
\begin{equation}
    \xi=2\tan^{-1}\left(\frac{r_0}{r}\right)
\end{equation}
where \(r_0\) is an arbitrary scaling coming from integration constant. This BPS skyrmion represent another class of \(D\)-brane soliton, firstly proposed in \cite{Dolan:1987kg} which is constructed from a generalized Nambu-Goto action and is expected to be scale invariant. This fact can be easily seen from the energy functional (or static part of Lagrangian) which is invariant under spatial coordinate transformation given by \(\Vec{x}\rightarrow \mu\Vec{x}\).

The resulting BPS bound in this model given in \eqref{BPSbound2}, again, proves that in any arbitrary \(d\) we always have a \(D\)-brane soliton (or specifically, skyrmion) which saturates the linear energy, but in a quite different manner from the previous submodel above. Firstly, we could easily notice that the Lagrangian does not obey the symmetry under all volume-preserving diffeomorphisms on \(S^d\) unless the self-duality condition is satisfied. For the case of hedgehog ansatz, this can only be satisfied for \(B=1\) skyrmion, implying a spherical symmetry. Several studies have been done in \(B>1\) hedgehog setups and it is found that \(B>1\) cases cannot saturate the bound \eqref{BPSbound2}, resulting in a repulsive configuration that cannot be stable \cite{Fadhilla:2020rig}. Secondly, in this submodel there is no potential term to break the symmetry, thus the target space symmetry of \(d\) dimensional sphere is conserved, provided that self-duality conditions are satisfied. It is also interesting that in hedgehog ansatz, the resulting skyrmion profile does not depend on spatial dimensions \(d\) which implies perfectly identical solution and features of skyrmion. 

By substituting the Bogomolnyi equation \eqref{Bog2} to \eqref{EnMomTen2} we arrive at the following results
\begin{equation}
    T_{00}=\varepsilon,~T_{0j}=T_{ij}=0~\text{for all values of}~i,j~\text{except 0,}
\end{equation}
where \(\varepsilon\) is the energy density of this second submodel, given by
\begin{equation}
    \varepsilon=\gamma_{d/2}C^d_{d/2}2^d\left(\frac{r_0}{r_0^2+r^2}\right)^d~.
\end{equation}
The finiteness of total energy is guaranteed because in this submodel we do not have a potential term. We can observe that the energy density takes a non-zero value at \(r=0\), namely
\begin{equation}
    \varepsilon(r=0)=2^dC^d_{d/2}\frac{\gamma_{d/2}}{r_0^d}
\end{equation}
which depends only on the scaling \(r_0\) and coupling constant \(\gamma_{d/2}\). This property implies that for \(B=1\) skyrmions are distributed close to the coordinate origin. Because dependency of \(\gamma_{d/2}\) can be easily omitted by scaling of the coordinate on the Lagrangian level, hence we are left with only \(r_0\) as the only arbitrary constant\footnote{Usually, the integration constants of Bogomolnyi equation is not arbitrary due to the specific boundary conditions, but in this special submodel the boundary conditions are satisfied for all size modulus \(r_0\in\mathbb{R}\) which is the consequence of scale invariance.}. 

\section{Conclusions and Outlooks}\label{sect:concl}
We have considered the higher dimensional Skyrme model and deduced that we can make a slight modification on generalized hedgehog ansatz by introducing winding numbers \(n_1\) to \(n_N\) for every azimuthal coordinate to accommodate higher values of topological degree. The winding numbers themselves are non-topological but the product of all winding numbers is proportional to the topological degree of the model. This form of ansatz is an extension of similar  forms found in \(d=2\) and \(d=4\) skyrmions \cite{Piette:1994mh,Brihaye:2017wqa}. 
In this work, we consider only the static case and we are going to address the more restrictive time-dependent cases in future works.

We have shown via BPS Lagrangian method that, in general, there exist at most two independent BPS Skyrme submodels within the static Skyrme field ansatz. The first BPS submodel is the model with only \(\mathcal{L}_{2d}\) term and potential term which can be found in any \(d\in\mathbb{N}\) dimensional spatial manifold. The second BPS submodel is the model with only \(\mathcal{L}_d\) term which can only be found in an even-dimensional spatial manifold. Both have been proposed by different hierarchies in \cite{Arthur:1996ia}.

The first BPS submodel has both compacton and regular solutions for the pion-mass potential to the power of \(s\) and lowers the bound of energy which is proportional to the topological degree as expected. This submodel is the higher dimensional extension of the known BPS Skyrme model in four and five spacetime dimensions \cite{Adam:2010zz, Fadhilla:2020rig}. The profile of the solution of this submodel highly depends on the form of potential and the dimensions of the target space. We found that the total energy is always finite for the pion-mass potential to the power of \(s\) with arbitrary power \(s\), and both compacton and regular solutions are non-interacting if the BPS bound is saturated.

The second submodel only has a regular solution which is similar to the one found in the five-dimensional spacetime case \cite{Brihaye:2017wqa, Fadhilla:2020rig}. The resulting Bogomolnyi equation is similar for every value of \(d\) up to a constant which depends on both \(d\) and the configuration of winding numbers. In contrast with the Bogomolnyi equation, the energy density has a distinctive feature that depends only on the topological degree. The energy density of the second submodel is concentrated at the origin for \(B=1\). Other important features of this type of BPS Skyrmion are that the solution and energy functional of this submodel are scale-invariant and BPS limit saturation happens if and only if all strain tensor eigenvalues are identical, i.e.
\begin{equation}
    \lambda_1^2=\lambda_2^2=\dots=\lambda_d^2~.
\end{equation}
This is the self-duality condition in terms of strain tensor eigenvalues.

In a more general case where a wider set of symmetry is considered with a higher topological degree, we expect that BPS features of the first submodel are maintained because the key term, which is the \(\mathcal{L}_{2d}\) term, is just a product of all eigenvalues. This is further supported by the expression of the energy which can only be altered if we choose another kind of compact manifold other than \(S^d\) as the target space. In contrast with the first one, the second submodel needs to satisfy a strong condition where all eigenvalues are the same. This might lead to smaller sets of possible target spaces where BPS solutions exist for this submodel. It is interesting to point out some, or all, target space symmetries that are compatible with this second submodel, this topic will be addressed in future works.

There exists non-self-dual solutions of the second submodel for higher topological degree but these solutions are found to be non-BPS since they do not saturate the BPS bound given in \eqref{BPSbound2}. Some aspects of these \(B>1\) Skyrmions will be discussed elsewhere.

\appendix
\section{On The Polyspherical Coordinate}\label{polyspherical}
As we know that a standard unit \(d-1\) dimensional sphere manifold is defined as 
\begin{equation}
    S^{d-1}=\{\Vec{x}\in \mathbb{R}^{d}, \|\Vec{x}\|=1\},
\end{equation}
where \(\mathbb{R}^{d}\) is a \(d\) dimensional Euclidean manifold and \(\|\Vec{x}\|\) is the Euclidean norm of \(\mathbb{R}^{d}\), defined as
\begin{equation}
    \|\Vec{x}\|\equiv \sqrt{\sum_{i=1}^d \left(x^i\right)^2}.
\end{equation}
We define a rotational plane to be a \(\mathbb{R}^2=\mathbb{R}^+\times S^1\) manifold equipped with coordinate system \(\{\rho,\varphi\}\) and metric \(ds_2^2=d\rho^2+\rho^2d\varphi^2\). Thus, a \(S^{d-1}\) has \(N=\left\lfloor\frac{d}{2}\right\rfloor\) independent rotational planes due to the fact that we can decompose \(\mathbb{R}^d\) into a product of some \(\mathbb{R}^2\)s. For even \(d\) the decomposition is given by
\begin{equation}\label{DecEven}
    \mathbb{R}^d=\underbrace{\mathbb{R}^2\times\mathbb{R}^2\times\dots\times\mathbb{R}^2}_{N=\frac{d}{2}},
\end{equation}
and for odd \(d\) we have the following decomposition
\begin{equation}\label{DecOdd}
    \mathbb{R}^d=\underbrace{\mathbb{R}^2\times\mathbb{R}^2\times\dots\times\mathbb{R}^2}_{N=\frac{d-1}{2}}\times\mathbb{R}.
\end{equation}
In a conclusion, we have been able to parametrize a \(d\) dimensional Euclidean manifold into patches of rotational plane coordinates \(ds_2^2=d\rho^2+\rho^2d\varphi^2\). Let us explain the details of this parametrization for both odd and even \(d\) cases in the subsections below.
\subsection{Even \(d\) Case}
From the decomposition given in \eqref{DecEven} we have the following metric for even \(d\) case,
\begin{eqnarray}
ds^2=\sum_{i=1}^d \left(dx^i\right)^2=\sum_{i=1}^N \left(d\rho_i^2+\rho_i^2d\varphi_i^2\right).
\end{eqnarray}
with \(N=\frac{d}{2}\).
Suppose that we have \(\mu_i\) such that, \(\rho_i\equiv r\mu_i\), relating the radial coordinate of one rotational plane to the radial coordinate of full space, then we have the following relation
\begin{eqnarray}
d\rho_i^2+\rho_i^2d\varphi_i^2=\mu_i^2dr^2+r^2d\mu_i^2+r^2\mu_i^2d\varphi_i^2+2r\mu_idrd\mu_i,
\end{eqnarray}
with no summation for repeated indices. Recall, that \(r=\|\Vec{x}\|=\sqrt{\sum_{i=0}^N\rho_i^2}=\sqrt{r^2\sum_{i=0}^N\mu_i^2}\), hence, we need to impose \(\sum_{i=1}^N\mu_i^2=1\) which leads to \(\sum_{i=0}^N\mu_id\mu_i=0\). Substituting these properties of \(\mu_i\) to the metric we have
\begin{eqnarray}
ds^2&=&\sum_{i=1}^N \left(d\rho_i^2+\rho_i^2d\varphi_i^2\right)\nonumber\\
&=&dr^2\sum_{i=1}^N\mu_i^2+r^2\sum_{i=1}^Nd\mu_i^2+r^2\sum_{i=1}^N\mu_i^2d\varphi_i^2+2rdr\sum_{i=1}^N\mu_id\mu_i\nonumber\\
&=&dr^2+r^2\left(\sum_{i=1}^Nd\mu_i^2+\sum_{i=1}^N\mu_i^2d\varphi_i^2\right).
\end{eqnarray}
This is the desired form of metric where we have \(N\) azimuthal coordinates, \(\varphi_i\), each of them living in a rotational plane.

It is conventional to take \(r\geq 0\) and \(0\leq \varphi_i < 2\pi\) for every \(i\). Thus, we only need to specify the form of \(\mu_i\) in terms of polar coordinates, \(\theta_i\), that represent the relative embedding position of rotational planes in the full space, \(\mathbb{R}^d\). One of possible parametrization for \(\mu_i\) is given by
\begin{eqnarray}
    \mu_1&=&\cos\theta_1~,\nonumber\\
    \mu_2&=&\sin\theta_1\cos\theta_2~,\nonumber\\
    &&\vdots\nonumber\\
    \mu_{N-1}&=&\sin\theta_1\dots\cos\theta_{N-1}~,\nonumber\\
    \mu_{N}&=&\sin\theta_1\dots\sin\theta_{N-1}~.
\end{eqnarray}
From the definition, we know that \(\mu_i=\rho_i/r\) and \(0\leq\rho_i\leq r\). Thus, \(0\leq \mu_i \leq 1\) which means that every polar coordinate must satisfy \(0\leq\theta_i\leq \frac{\pi}{2}\). This completes our parametrization of \(S^{d-1}\) in polyspherical coordinate for even \(d\) case.

Obtaining the inverse transformation to the standard Euclidean coordinate \(x^i\) is straightforward. Firstly, we need to divide the \(x^i\)s into two sets with one set contains coordinate of every horizontal axes of submanifold \(\mathbb{R}^2\) and the other set contains the vertical axes coordinate. Thus, we have the following relations
\begin{eqnarray}\label{EucCoorEveni}
    x^i&=&r\mu_i\cos\varphi_i,\\
    x^{2i}&=&r\mu_i\sin\varphi_i.\label{EucCoorEven2i}
\end{eqnarray}
where \(i\) runs from \(1\) to \(N\).
\subsection{Odd \(d\) Case}
We can apply the same method from previous subsection for decomposition \eqref{DecOdd} with extra coordinate \(\rho_0\) for submanifold \(\mathbb{R}\) such that the metric for odd \(d\) case is given by
\begin{eqnarray}
ds^2=\sum_{i=1}^d \left(dx^i\right)^2=d\rho_0^2+\sum_{i=1}^N \left(d\rho_i^2+\rho_i^2d\varphi_i^2\right).
\end{eqnarray}
with \(N=\frac{d-1}{2}\). The constraint for \(\mu_i\) now becomes \(\mu_0^2+\sum_{i=1}^N\mu_i^2=1\), which leads to \(\mu_0d\mu_0+\sum_{i=1}^N\mu_id\mu_i=0\). Again, by substituting the constraint for \(\mu_i\)s and its implications to the metric, we arrive at the following expression
\begin{eqnarray}
ds^2&=&r^2d\mu_0^2+2r\mu_0drd\mu_0+\mu_0^2dr^2+\sum_{i=1}^N \left(d\rho_i^2+\rho_i^2d\varphi_i^2\right)\nonumber\\
&=&dr^2\left(\mu_0^2+\sum_{i=1}^N\mu_i^2\right)+r^2\left(d\mu_0^2+\sum_{i=1}^Nd\mu_i^2\right)+r^2\sum_{i=1}^N\mu_i^2d\varphi_i^2+2rdr\left(\mu_0d\mu_0+\sum_{i=1}^N\mu_id\mu_i\right)\nonumber\\
&=&dr^2+r^2d\mu_0^2+r^2\left(\sum_{i=1}^Nd\mu_i^2+\sum_{i=1}^N\mu_i^2d\varphi_i^2\right).
\end{eqnarray}
We can adopt a similar form of \(\mu_i\) from even \(d\) case with additional \(\mu_0\) term that gives us the following result
\begin{eqnarray}
    \mu_0&=&\cos\theta_0~,\nonumber\\
    \mu_1&=&\sin\theta_0\cos\theta_1~,\nonumber\\
    \mu_2&=&\sin\theta_0\sin\theta_1\cos\theta_2~,\nonumber\\
    &&\vdots\nonumber\\
    \mu_{N-1}&=&\sin\theta_0\sin\theta_1\dots\cos\theta_{N-1}~,\nonumber\\
    \mu_{N}&=&\sin\theta_0\sin\theta_1\dots\sin\theta_{N-1}~.
\end{eqnarray}
Consider, \(\mu_0=\frac{\rho_0}{r}\) and \(-r\leq \rho_0\leq r\). This leads to \(-1\leq \mu_0\leq 1\) which means that \(0\leq\theta_0\leq \pi\). The domain of the rest of the \(\theta_i\)s have been specified in previous subsection and that completes our parametrization of \(S^{d-1}\) in polyspherical coordinate for odd \(d\) case.

The inverse transformation to standard Euclidean coordinate for this case is similar with the one we found for even \(d\) case, namely
\begin{eqnarray}\label{EucCoorOddi}
    x^i&=&r\mu_i\cos\varphi_i,\\
    x^{2i}&=&r\mu_i\sin\varphi_i,\label{EucCoorOdd2i}\\
    x^d&=&r\mu_0.\label{mu0}
\end{eqnarray}
where \(i\) runs from \(1\) to \(N\). The only modification we need to accommodate \(\mu_0\) is the last equation \eqref{mu0} and we can choose to place at the \(d\)-th coordinate, \(x^d\).
\section{Bogomolnyi equation from Equation of Motions}
In this part of the paper we are going to show explicitly how we can reproduce the same differential equation of field \(\xi(r)\) from both BPS Lagrangian method and dynamical field equation \eqref{secondFieldEq}
\subsection{Field Equation of The First BPS Submodel}
Firstly, let us focus on the field equation of the first submodel which is recast in this following form
\begin{eqnarray}\label{dynEq1}
\left(\delta^{cb}-\phi^c\phi^b\right)\left[ \frac{2d\gamma_d}{d!}\nabla_{j}\left(\phi^{b,i}H_i^j(d,d)\right)-\frac{\partial}{\partial \phi^b} \gamma_0 V\right]=0~.
\end{eqnarray}
From the \(O(d+1\) model constraint, we have \(\phi^a\phi^a=1\) which implies that 
\begin{eqnarray}\label{impEq1}
\phi^a\phi^a_i=0~.
\end{eqnarray}
From this equation, we can prove another useful identity from the derivative of \(\phi^a\phi^a_i\), given by \(\nabla_j(\phi^a\phi^a_i)=\phi^a_i\phi^a_j+\phi^a\nabla_j\phi^a_i\), which leads to
\begin{equation}\label{impEq2}
    \phi^a\nabla_j\phi^a_i=-\phi^a_i\phi^a_j=-D_{ij}~.
\end{equation}
Applying these two identities to \eqref{dynEq1} gives
\begin{eqnarray}
    \frac{2\gamma_d}{(d-1)!}\left[\nabla_{j}\left(\phi^{c,i}H_i^j(d,d)\right)+\phi^cH_i^j(d,d)D^i_j\right]-\gamma_0\left[\frac{\partial}{\partial \phi^c} V-\phi^c\phi^b\frac{\partial}{\partial \phi^b} V\right]=0~.
\end{eqnarray}
Recall that we choose to work specifically on potential which takes the form \(V\equiv V(\xi)\) and \(\phi^a\phi^a=1\) where the only field which does not contain angular factor is \(\phi^0=\cos\xi\), then the summation \(\phi^b\frac{\partial}{\partial \phi^b} V\) is equal to \(\phi^0\frac{\partial}{\partial \phi^0} V(\xi)=-V'(\xi)\cot\xi \) because the other term for \(b\neq 0\) vanishes. Upon substitution of the ansatz to equations \eqref{dynEq1} we arrive at a single equation for \(\xi\), namely
\begin{equation}\label{FielEqSubmodFirst}
    \frac{2\gamma_d}{(d-1)!}\left[-\frac{1}{r^{d-1}}\frac{d}{dr}\left(r^{d-1}\sin\xi\xi'\left(B^2\frac{\sin^{2(d-1)}\xi}{r^{2(d-1)}}\right)\right)+d!\cos\xi B^2\frac{\sin^{2(d-1)}\xi}{r^{2(d-1)}}\xi'{}^2\right]+\gamma_0V'(\xi)\sin\xi =0
\end{equation}
and all other equations only differ by some angular factors. Equation \eqref{FielEqSubmodFirst} is equivalent to Euler-Lagrange equations derived from effective Lagrangian. With some manipulations on field equation \eqref{FielEqSubmodFirst} we can simplify it to become
\begin{equation}
    B\frac{ \sin^{d-1}\xi}{r^{d-1}}\frac{d}{dr}\left(B\frac{\sin^{d-1}\xi}{r^{d-1}}\xi'\right)=\frac{\gamma_0}{2\gamma_d}V'(\xi)~.
\end{equation}
Multiplying both sides with \(\xi'\) gives
\begin{equation}
    \frac{1}{2}\frac{d}{dr}\left(B\frac{\sin^{d-1}\xi}{r^{d-1}}\xi'\right)^2=\frac{\gamma_0}{2\gamma_d}\frac{d}{dr}V(\xi)~,
\end{equation}
which implies a Bogomolnyi equation of the form
\begin{equation}
    B\frac{\sin^{d-1}\xi}{r^{d-1}}\xi'\mp \sqrt{\frac{\gamma_0}{\gamma_d}}\sqrt{V(\xi)}=0~.
\end{equation}
As a conclusion, it is proven that we can rederive the Bogomolnyi equation of the first submodel from its full field equation, implying a consistent formulation of BPS Lagrangian method.
\subsection{Field Equation of The Second BPS Submodel}\label{AppB2}
Again, let us consider the field equation for this submodel, given by
\begin{eqnarray}
\left(\delta^{cb}-\phi^c\phi^b\right) \frac{d\gamma_{d/2}}{\left((d/2)!\right)^2}\nabla_{j}\left(\phi^{b,i}H_i^j(d/2,d)\right)=0~.
\end{eqnarray}
By exploiting the identity given in \eqref{impEq1} we can simplify the equation above to become
\begin{equation}\label{dynEq2}
    \nabla_{j}\left(\phi^{c,i}H_i^j(d/2,d)\right)-\phi^cH_i^j(d/2,d)\nabla_{j}\phi^{b,i}=0~.
\end{equation}
Now consider only the second term of the equation above. From equation \eqref{impEq2} we can recast it into the following form
\begin{equation}
    -H_i^j(d/2,d)\nabla_{j}\phi^{b,i}=H_i^j(d/2,d)D^i_j=\left((d/2)!\right)^2\left({}_dK_{(d/2)-1}\xi'^2+{}_dK_{d/2}\frac{\sin^2\xi}{r^2}\right)\frac{\sin^{d-2}\xi}{r^{d-2}}~.
\end{equation}
If we substitute the ansatz into \eqref{dynEq2}, then we arrive at \(d+1\) equations which are equivalent and differs only by some angular factor. Thus, the equations become, effectively, single equation of \(\xi\), given by
\begin{eqnarray}\label{FielEqSubmodScnd}
    &&-\frac{1}{r^{d-1}}\frac{d}{dr}\left(r^{d-1}\sin\xi\xi'\left((d/2)!{}_dK_{(d/2)-1}\frac{\sin^{d-2}\xi}{r^{d-2}}\right)\right)\nonumber\\&&+\cos\xi\left((d/2)!\right)^2\left({}_dK_{(d/2)-1}\xi'^2+{}_dK_{d/2}\frac{\sin^2\xi}{r^2}\right)\frac{\sin^{d-2}\xi}{r^{d-2}}=0~.
\end{eqnarray}
Equation \eqref{FielEqSubmodScnd} is equivalent to the Euler-Lagrange equation derived from the effective Lagrangian. Now, Suppose that the corresponding Bogomolnyi equation of this second submodel is given by
\begin{equation}
    \xi'=-\alpha\frac{\sin\xi}{r}~,
\end{equation}
then, by substituting this expression into \eqref{FielEqSubmodScnd} we arrive at
\begin{equation}
    \frac{2^dd}{r}\left(-1+\frac{r_0^{2\alpha}}{r^{2\alpha}}\right)\left(\frac{1}{\frac{r^a}{r_0^a}+\frac{r_0^a}{r^a}}\right)\left(\alpha^2{}_dK_{(d/2)-1}-{}_dK_{d/2}\right)=0~.
\end{equation}
Since the equation above must be satisfied for all \(r\) then we need to take \(\left(\alpha^2{}_dK_{(d/2)-1}-{}_dK_{d/2}\right)=0\) which lead to \(\alpha=\sqrt{\frac{{}_dK_{d/2}}{{}_dK_{(d/2)-1}}}\). We can observe that we arrive at the same expression for \(\alpha\) from BPS Lagrangian method, thus we can conclude that the ansatz is consistent for both Bogmolnyi equation and full field equation from Euler-Lagrange equation.
\section{Components of Energy-Momentum Tensor via Eigenvalues of Strain Tensor}
The strain tensor given in \eqref{Strain} possesses a set of eigenvalues of given by \eqref{EigvalEven} for even \(d\) and \eqref{EigvalOdd} for odd \(d\). These simplified version of eigenvalues actually came from 
\begin{equation}\label{Eigvalfull}
    \lambda_1^2=g^{11}\xi'^2,~~~\text{and}~\lambda_k^2=g^{kk}\sin^2\xi \partial_k \Vec{n}\cdot\partial_k \Vec{n}~\text{for}~k=2,\dots,d~.
\end{equation}
with no summation on repeating indices.
In equation \eqref{Eigvalfull} we can see the reason why choosing a suitable ansatz for a particular coordinate is important to reduce the complexity of the resulting effective model. The main rule for the choice of ansatz which we have used above is that we need the ansatz to satisfy proportionality \(\partial_k \Vec{n}\cdot\partial_k \Vec{n}\propto g_{kk}\), such that the quantity \(g^{kk}\partial_k \Vec{n}\cdot\partial_k \Vec{n}\) does not depend on any angular coordinates.

We can exploit the eigenvalues given in \eqref{Eigvalfull} to calculate the energy-momentum tensor more simply. Firstly, recall that the energy-momentum tensor is given by
\begin{equation}
    T_{ij}=-2\frac{\partial \mathcal{L}}{\partial g^{ij}}+g_{ij}\mathcal{L},
\end{equation}
and Skyrme model Lagrangian expressed in terms of eigenvalues is given by
\begin{eqnarray}
    \mathcal{L}&=&-\left[\gamma_0 V+\sum_{i=1}^d \gamma_i e_i(\lambda_1^2,\dots,\lambda_d^2)\right]\nonumber\\
    &=&-\left[\gamma_0 V+\sum_{i=1}^d \gamma_i e_i(g^{11}\xi'^2,g^{22}\sin^2\xi,\dots,n_N^2 g^{dd}\sin^2\xi \sin^2\theta_0\dots\sin^2\theta_{N-1})\right].
\end{eqnarray}
Because all of the arguments of symmetric polynomials \(e_i\) contains \(g^{kk}\), then the term \(\frac{\partial \mathcal{L}}{\partial g^{ij}}\) implicitly removes all the term which are not proportional to \(g^{ij}\). It is easy to prove that all the \(T_{ij}\) for \(i\neq j\) vanishes, hence there only \(T_{kk}\) components are left. This \(T_{kk}\) components can be easily calculated as the full effective Lagrangian subtracted by two times the terms which are proportional to \(g^{kk}\) which means that the eigenvalues which are proportional to \(g^{kk}\), \(\lambda_k^2\), changes their sign to \((-)\). In other words, we have the following form of non-vanishing components of energy-momentum tensor, given by
\begin{equation}\label{EnMomTenAnsatz}
    T_{kk}=-g_{kk}\left[\gamma_0 V+\sum_{i=1}^d \gamma_i e_i(\lambda_1^2,\dots,-\lambda_k^2,\dots,\lambda_d^2)\right]~.
\end{equation}
It is important to note that the expression of the energy-momentum tensor in \eqref{EnMomTenAnsatz} might not be the most general one, since we restrict our case to the ansatz satisfying the properties mentioned above to suit the spacetime metric.

Let us focus on the case of BPS submodels. The spatial diagonal component of the energy-momentum tensor for the first submodel is given by
\begin{equation}
    T^{(1)}_{kk}=-g_{kk}\left[\gamma_0 V+ \gamma_d e_d(\lambda_1^2,\dots,-\lambda_k^2,\dots,\lambda_d^2)\right]=-g_{kk}\left[\gamma_0 V- \gamma_d \text{det}D\right]~.
\end{equation}
Substitution of the Bogomolnyi equation, \(\sqrt{\gamma_0 V}\pm \sqrt{\gamma_d \text{det}D}=0\) to \(T^{(1)}_{kk}\) above gives
\begin{equation}
    T^{(1)}_{kk}=0~\text{for}~k=1,\dots,d~.
\end{equation}
Thus, for every \(B\) the spatial components of energy-momentum tensor of the first submodel are equal to zero. For the second submodel, the spatial diagonal components of energy-momentum tensor is given by
\begin{equation}
    T^{(2)}_{kk}=-g_{kk}\gamma_{d/2}~ e_{d/2}(\lambda_1^2,\dots,-\lambda_k^2,\dots,\lambda_d^2)~.
\end{equation}
If we impose the self-duality condition for the second submodel where all eigenvalues are identical, then the equation becomes \(T^{(2)}_{kk}=-g_{kk}\gamma_{d/2}\lambda_k^2\left(C^{d-1}_{d/2}-C^{d-1}_{(d/2)-1}\right)=0\), which implies that, if BPS limit is saturated then we always have vanishing spatial components of the energy-momentum tensor. This is not the case for non-self-dual Skyrmion with \(B>1\), implying an unstable Skyrmion solution. We demonstrate this property as follows:
Suppose that we take the first eigenvalues to be \(\xi'{}^2\) and factorize it from \(e_{d/2}\) or \(e_{N}\),\footnote{we take \(\frac{d}{2}=N\) for this special case from now on to simplify notations.} then we arrive at the following relation
\begin{eqnarray}
    T^{(2)}_{kk}&=&-g_{kk}\gamma_{N}e_{N}(\lambda_1^2,\dots,-\lambda_k^2,\dots,\lambda_d^2)\nonumber\\&=&-g_{kk}\gamma_{N}\left[\lambda_1^2e_{N-1}(\lambda_2^2,\dots,-\lambda_k^2,\dots,\lambda_d^2)+e_{N}(\lambda_2^2,\dots,-\lambda_k^2,\dots,\lambda_d^2)\right]~.
\end{eqnarray}
We know that the rest of the eigenvalues are proportional to \(\frac{\sin^2\xi}{r^2}\), hence it is useful to factorize it and then divide the case of energy-momentum tensor into polar components corresponding to \(\theta_k\) and azimuthal components corresponding to \(\varphi_k\). Since elementary symmetric polynomials satisfies the following identity
\begin{equation}\label{identESP}
    e_n(x_1,\dots,x_m)=\sum_{i=0}^ne_{n-i}(x_1,\dots,x_k)e_i(x_{k+1},\dots,x_m)~,
\end{equation}
then we can divide the eigenvalues related to \(N-1\) polar coordinates which are proportional to 1 and the eigenvalues related to \(N\) azimuthal coordinates which are proportional to their corresponding winding numbers \(n_k\) and substitute it into \eqref{identESP} after factorizing the \(\frac{\sin^2\xi}{r^2}\) factor. This process lead us to the following form of polar components
\begin{eqnarray}\label{nonvanishTpol}
    T^{(2)}_{22}&=&-r^2\xi'{}^2\frac{\sin^{d-2}\xi}{r^{d-2}}\sum_{i=0}^{N-1}e_{N-1-i}(\underbrace{1,\dots,-1,\dots,1}_{N-1})e_{i}(n_1^2,\dots,n_N^2)\nonumber\\
    &&-r^2\frac{\sin^{d}\xi}{r^{d}}\sum_{i=0}^{N}e_{N-i}(\underbrace{1,\dots,-1,\dots,1}_{N-1})e_{i}(n_1^2,\dots,n_N^2)\nonumber\\
    &=&-r^2\xi'{}^2\frac{\sin^{d-2}\xi}{r^{d-2}}\sum_{i=0}^{N-1}\left(C^{N-2}_{N-1-i}-C^{N-2}_{N-2-i}\right)e_{i}(n_1^2,\dots,n_N^2)\nonumber\\&&-r^2\frac{\sin^{d}\xi}{r^{d}}\sum_{i=0}^{N}\left(C^{N-2}_{N-i}-C^{N-2}_{N-1-i}\right)e_{i}(n_1^2,\dots,n_N^2)~,\\
    g^{22}T^{(2)}_{22}&=&g^{33}T^{(2)}_{33}=g^{44}T^{(2)}_{44}=\dots=g^{NN}T^{(2)}_{NN}~,
\end{eqnarray}
and for azimuthal components we have 
\begin{eqnarray}\label{nonvanishTazi}
    T^{(2)}_{(N+k)(N+k)}&=&-g_{(N+k)(N+k)}\xi'{}^2\frac{\sin^{d-2}\xi}{r^{d-2}}\sum_{i=0}^{N-1}e_{N-1-i}(\underbrace{1,\dots,1}_{N-1})e_{i}(n_1^2,\dots,-n_k^2,\dots,n_N^2)\nonumber\\
    &&-g_{(N+k)(N+k)}\frac{\sin^{d}\xi}{r^{d}}\sum_{i=0}^{N}e_{N-i}(\underbrace{1,\dots,1}_{N-1})e_{i}(n_1^2,\dots,-n_k^2,\dots,n_N^2)\nonumber\\
    &=&-g_{(N+k)(N+k)}\xi'{}^2\frac{\sin^{d-2}\xi}{r^{d-2}}\sum_{i=0}^{N-1}C^{N-1}_{N-1-i}e_{i}(n_1^2,\dots,-n_k^2,\dots,n_N^2)\nonumber\\&&-g_{(N+k)(N+k)}\frac{\sin^{d}\xi}{r^{d}}\sum_{i=0}^{N}C^{N-1}_{N-i}e_{i}(n_1^2,\dots,-n_k^2,\dots,n_N^2)~.
\end{eqnarray}
We can observe that all these spatial components of \(B>1\) cases are non-zero, but unfortunately equation \eqref{nonvanishTpol} stops to work for \(d=2\) where \(N=1\). This peculiarity leads to vanishing spatial components as we show below.

Let us consider the case of \(d=2\) which has two eigenvalues,
\begin{equation}
    \lambda_1^2=\xi'^2,~~~\lambda_2^2=n_1^2\frac{\sin^2\xi}{r^2}~.
\end{equation}
The \(T_{11}\) and \(T_{22}\) components are given by
\begin{eqnarray}\label{}
    T_{11}&=&-\left[\gamma_0 V+\gamma_1\left(-\xi'^2+n_1^2\frac{\sin^2\xi}{r^2}\right)-\gamma_2\left(n_1^2\xi'^2\frac{\sin^2\xi}{r^2}\right)\right]~,\\
    T_{22}&=&-r^2\left[\gamma_0 V+\gamma_1\left(\xi'^2-n_1^2\frac{\sin^2\xi}{r^2}\right)-\gamma_2\left(n_1^2\xi'^2\frac{\sin^2\xi}{r^2}\right)\right]~.
\end{eqnarray}
We can directly see if we consider the first BPS submodel (\(\gamma_1=0\)) and impose the BPS limit then both \(T_{11}\) and \(T_{22}\) vanish. This property can be found in all first BPS submodels for arbitrary \(d\). In contrast with the first submodel, we can observe that the second BPS submodel (\(\gamma_0=\gamma_2=0\)) does have vanishing angular components of energy-momentum tensor for the case of \(d=2\) but we have shown in \eqref{nonvanishTpol} and \eqref{nonvanishTazi} that they are non-zero for \(d>2\). This property leads to a stable self-dual BPS Skyrmion within this ansatz for the second BPS submodel which generally cannot be found for \(d\) other than 2.
\acknowledgments
The work in this paper is supported by GTA Research Group ITB, Riset ITB 2022, and PDUPT Kemenristek-ITB 2022. A. N. A. would like to acknowledge the support from the ICTP through the Associates Programme (2018-2023). B. E. G. would also acknowledge the support from the ICTP through the Associates Programme (2017-2022). E. S. F. also would like to acknowledge the support from BRIN through the Research Assistant Programme 2022.

\medskip
$\bibliographystyle{utphys}$
\bibliography{referenceER.bib}

\end{document}